\documentclass[amssymb,aps,twocolumn,showpacs, superscriptaddress,nofootinbib]{revtex4-1}
\usepackage[]{graphicx}
\usepackage[]{amsmath}
\usepackage{hyperref}

\def\ket#1{| #1 \rangle}

\def\cO{\mathcal{O}}

\def\cT{\mathcal{T}}

\def\eq#1{Eq.~\eqref{eq:#1}}

\def\one{{\mathchoice {\rm 1\mskip-4mu l} {\rm 1\mskip-4mu l} {\rm
1\mskip-4.5mu l} {\rm 1\mskip-5mu l}}}

\begin{document}

\title{Quantum simulation of time-dependent Hamiltonians and the convenient illusion of Hilbert space}
\author{David Poulin}
\affiliation{D\'epartement de Physique, Universit\'e de Sherbrooke, Qu\'ebec, Canada}
\author{Angie Qarry}
\affiliation{Faculty of Physics, University of Vienna Boltzmanngasse
5, 1090 Vienna, Austria}
\affiliation{Centre for Quantum Technologies, National University of Singapore, Singapore 117543, Singapore}
\author{R. D. Somma}
\affiliation{Los Alamos National Laboratory, Los Alamos, NM 87545, USA}
\author{Frank Verstraete}
\affiliation{Faculty of Physics, University of Vienna Boltzmanngasse
5, 1090 Vienna, Austria}

\date{\today}

\begin{abstract}
We consider the manifold of all quantum many-body states that can be
generated by arbitrary time-dependent local Hamiltonians in
a time that scales polynomially in the system size, and show that it
occupies an exponentially small volume in Hilbert space. This
implies that the overwhelming majority of states in Hilbert space are not physical as they can only be produced after an exponentially long time. We establish this fact by making use of a time-dependent generalization of the Suzuki-Trotter expansion, followed by a counting argument. This also demonstrates that a computational model based on arbitrarily rapidly changing Hamiltonians is no more powerful than the standard quantum circuit model.
\end{abstract}


\maketitle

The Hilbert space of a quantum system is big---its dimension grows exponentially with the number of particles it contains. Thus, parametrizing a generic quantum state of $N$ particles requires an exponential number of real parameters. Fortunately, the states of many physical systems of interest appear to occupy a tiny sub-manifold of this gigantic space. Indeed, the essential physical features of many systems can be explained by variational states specified with a small number of parameters.  Well known examples include the BCS state for superconductivity \cite{BCS57a}, Laughlin's state for fractional quantum Hall liquids \cite{L83a}, tensor network states occurring in real-space renormalization methods \cite{CV09}. In these cases, the number of parameters scales only polynomially with $N$.

In this Letter, we attempt to define the class of {\em physical} states of a many-body
quantum system with local
Hilbert spaces of bounded dimensions, and prove that they represent an exponentially small sub-manifold of the Hilbert space. We say that a state is physical
if it can be reached, starting in some fiducial state (e.g. a ferromagnetic state, or the vacuum), by an evolution generated by any time-dependent
quantum many-body Hammiltonian, with the
constraint that 1) the Hamiltonian is local in the sense that it is the sum of terms each acting on at most $k$ bodies for some constant $k$ independent of $N$, and  2) the duration of the evolution scales at most as a polynomial in the number of particles in the system. The assumption about the initial fiducial state is artificial; we could alternatively define
the class of physical evolutions for quantum many-body systems as the ones  generated by Hamiltonians obeying constraints 1 and 2, and would reach the same conclusions.

The second constraint is very much reminiscent of the way complexity
classes are defined in theoretical computer science, where that central object of study is the scaling of the time required to solve a problem as a function of its input size.  The classical analogue for the problem that we address
is a well known counting argument of Shannon~\cite{S49a} demonstrating that the number of boolean functions of $N$ bits scales
doubly exponentially (as $2^{2^N}$), with the consequence that no efficient
(i.e. polynomial) algorithm can exist to compute the overwhelming majority of those
functions. Indeed, the number of different functions that can be encoded by
all classical circuits of polynomial depth scale as
$2^{poly(N)}$, which is exponentially smaller than the total number
of Boolean functions.

Our contribution is a quantum generalization of this result, which has in part been a folklore theorem in the quantum information community for some time. The crux of our argument is to demonstrate that the dynamics generated by any local Hamiltonian, without any assumptions on its time-dependence, can be simulated by a quantum circuit of polynomial size. All previously known simulation methods \cite{Llo96b,Zal98b,AT03b,BACS07a,BC09a1,CK10a1} produced a quantum circuit of complexity that depends on the smoothness of the Hamioltonian, scaling e.g. with $\|\partial H/\partial t\|$ or some higher derivatives. Using the results of Huyghebaert and De Raedt \cite{HD90a}, we show how this conditions can be overcome, and demonstrate how the counting argument of Shannon can be repeated in the quantum case by invoking the Solovay-Kitaev theorem \cite{Kitaev}.  Note that a direct parameter counting would not produce this result because we impose no restriction on the time-dependence of the Hamiltonian. The complete description of a rapidly changing Hamiltonian requires lots of information, so from this perspective there are in principle enough parameters to reach all states in the Hilbert space. This leads to the conclusion that most states in the Hilbert space are not physical: they can only be reached after an exponentially long time.  This has to be contrasted to the classical case, where all \emph{states} of $N$ bits
correspond to physical states: they can easily be generated by trivial depth-one circuits. The difference between classical and quantum behavior is due to the existence of quantum entanglement.

Our demonstration that arbitrary local time-dependent Hamiltonians can be efficiently simulated on quantum computers is of interest in its own right in the context of quantum computation.  More precisely, we are concerned with Hamiltonians acting on $N$ particles of the form
\begin{equation}
H(t) = \sum_{X \subset \{1,2,\ldots, N\}} H_X(t)
\label{eq:H}
\end{equation}
where $X$ labels subsets of the $N$ particles, each term has bounded norm $\|H_X(t)\| \leq E$, and each term acts on no more than $k$ particles, i.e., $H_X(t) = 0$ if $|X| > k$, and $k$ is fixed independent of the system size. We make no assumption on the geometry of the system and the coupling can be arbitrarily long ranged. The time-evolution operator $U(0,t)$ from time 0 to $t$ is governed by Schr\"odinger's equation $\frac d{dt}U(0,t) = -iH(t)U(0,t)$, with solution given in terms of a time-ordered integral $U(0,t) = \cT \exp\left\{ \int_0^t H(s) ds\right\}$.

Starting with Feynman's exploration of quantum computers \cite{Fey82a}, it has been well established that the time-evolution operator generated by Hamiltonians of the form \eq{H}  can be decomposed into short quantum circuits, provided that $H(t)$ varies slowly enough \cite{Llo96b,Zal98b,AT03b,BACS07a,BC09a1,CK10a1}. In all cases, this is achieved by approximating the evolution operator by a {\em product formula}
\begin{equation}
U(0,t) \approx \prod_{p=1}^{N_p} \exp\left\{-i H_{X_p}(t_p) \Delta t_p\right\},
\label{eq:product}
\end{equation}
where the sequences $X_p$, $t_p$, and $\Delta t_p$ are set by specific approximation schemes, such as the Trotter formula \cite{T59a} or the Lie-Suzuki-Trotter formula \cite{S93b}.
Because each term $H_X$ acts on at most $k$ particles, this last expression represents a sequence of $k$-particles gates. A standard quantum circuit is obtained by decomposing each of these $k$-body operator as a sequence of one- and two-qubit gates using the result of Solovay-Kitaev \cite{Kitaev,DN06a1}.

Perhaps the simplest  example of a product formula decomposition of $U(0,t)$ is given by
\begin{equation}
U(0,t) \approx \prod_{j=1}^n \prod_X \exp\left\{-i H_X(j\Delta t)\Delta t\right\}
\label{eq:simplePF}
\end{equation}
where the product over $X$ can be carried in any given order. This decomposition makes use of two approximations. First, the time-dependence of the $H(t)$ is ignored on time-scales lower than $\Delta t$: the Hamiltonian is approximated by a piece-wise constant function taking the values $H(j\Delta t)$ on the time-interval $[(j-1)\Delta t,j\Delta t]$. Second, each matrix exponential is decomposed using the Trotter formula $\exp\{-iH(t) \Delta t\} \approx \prod_X \exp\{-iH_X(t) \Delta t H_X\}$.  Clearly, the size $\Delta t$ of the time intervals must be shorter than the fluctuation time-scale of $H(t)$ for the first approximation to be valid, $\Delta t \ll \|\partial H/\partial t\|^{-1}$. Higher frequency fluctuations would therefore require breaking the time-evolution into shorter intervals, thus increasing the overall complexity of the simulation.

\medskip
\noindent\textit{Time-dependent Trotter-Suzuki expansion---}Somewhat surprisingly, it is possible to generalize the Trotter-Suzuki formula to
time-dependent Hamiltonians without compromising the error, where
the Hamiltonian may exhibit fluctuations much faster than the
time step $\Delta t$. We begin by breaking the total time evolution into short segments $U(0,T)) =  U(t_{n},t_{n}+\Delta t)\ldots  U(t_2,t_2+\Delta t) U(0,0+\Delta t)$ duration of time $\Delta t$
\[{U}(t_j,t_j+\Delta t)=\mathcal{T}\exp\left(-i\int_{t_j}^{t_j+\Delta t} ds
\sum_X{H}_X(s)\right).\]
In the simple case where the sum over $X$ contains only two terms, say $H_1$ and $H_2$, it has been shown \cite{HD90a} that the generalized Trotter-Suzuki expansion
\begin{eqnarray*}{U}^{{\rm TS}}(t_j,t_j+\Delta t) &=&
\mathcal{T}\exp\left(-i\int_{t_j}^{t_j+\Delta t} ds
{H}_1(s)\right)\\
&&\times \mathcal{T}\exp\left(-i\int_{t_j}^{t_j+\Delta t} ds
{H}_2(s)\right)\end{eqnarray*}
gives rise to an  error in terms of operator norm that is equal to
\[ \|U(t_j,t_j+\Delta t)-U^{\rm TS}(t_j,t_j+\delta t)\|\leq c_{12}(\Delta
t)^2\] with $c_{12}$ of the order of 1 and given by
\[c_{12}=\frac{1}{(\Delta t)^2}\int_{t_j}^{t_j+\Delta t} dv\int_{t_j}^v du \|[H_1(u),H_2(v)]\|.\]
which is upper bounded by $c_{max}^2/2$ with $c_{max}=\max_t\max_X \|H_X(t)\|$.
Note that this bound does not depend on the derivative of the Hamiltonian
(and is therefore also valid for non-analytic time-dependence). Note
also that the bound reduces to the usual Trotter error for the
time-independent case and is therefore equally strong, and that it can straightforwardly be generalized to higher order decompositions.

For our present application, the Hamiltonian is the sum of $L \in {\rm poly}( N)$ $k$-body terms, c.f. \eq{H}. We can therefore iterate the above procedure $\log_2(L)$ times; at the $n$'th iteration, there are $2^n$ terms, each of strength upper bounded by $c_{max}L/2^n$. The total error for approximating the exact time-evolution of the Hamiltonian with $L$ terms by a product of $L$ time-ordered terms is therefore
\[\frac{1}{2}c_{max}^2(\Delta t)^2 \sum_{m=1}^{\log_2 L} 2^m\left(\frac{L}{2^m}\right)^2\leq \frac{1}{2}c_{max}^2 L^2 (\Delta t)^2\]
which can be made arbitrary small by choosing a $\Delta T$ that scales as an inverse polynomial in the number of qubits. Approximating the time evolution operator over a total time $T$ with a product of $k$-local unitaries such that the total error is $\epsilon/2$ can therefore be achieved by choosing
\[\Delta t=\frac{\epsilon}{Tc_{max}^2 L^2}.\]
The total number $G$ of  $k$ -local quantum gates to achieve this accuracy is then equal to
\[G(\epsilon,L)=L\frac{T}{\Delta T}=\frac{c_{max}^2}{\epsilon}T^2L^3.\]

Each of these $k$-body unitaries can be decomposed into standard 2-qubit gates chosen from a discrete set of quantum gates (e.g. CNOT's between any pair of qubits supplemented by a local $\pi/8$ rotation gate) . The obvious approach is to exactly integrate the corresponding time-ordered exponential  on a classical computer, and then to use the Solovay-Kitaev theorem to count the number of standard quantum gates needed to approximate this. To achieve an accuracy $\epsilon_G$ per unitary, we need $d_{SK}\left(\log^{c_{SK}}(1/\epsilon_G)\right)$ standard gates with $c_{SK}$ and $d_{SK}$ constants; we choose $\epsilon_G$ such that  \[\epsilon_G=\frac{\epsilon}{2G(\epsilon,T)}.\] The total number of quantum gates as chosen from a discrete set of gates needed  to approximate the complete time evolution with an error $\epsilon$ is therefore upper bounded by
\[G_{tot}(\epsilon,T)=d_{SK}G(\epsilon,T)\log^{c_{SK}}\left(G(\epsilon,T)/\epsilon\right)\]
which is polynomial in the number of qubits; it roughy scales quadratically in $T$ and as the cube of the total number of non-commuting local terms in the Hamiltonian. This actually proves that quantum computers operating under arbitrary time-dependent local Hamiltonians are no more powerful than quantum computers based on the quantum circuit model.

\medskip
\noindent{\em Average Hamiltonians  and randomized evolution}---Note that the time-dependent Trotter-Suzuki decomposition described in the previous section does not lead to a product formula because each term appearing in it involves a time-ordered integral $U_{X}(t_j,t_j+\Delta t) = \cT \exp\{-i\int_{t_j}^{t_j+\Delta t} H_X(s)ds\}$ rather than the exponential of a term of the Hamiltonian at a given time $\exp\{-i\Delta t H_X(t)\}$ as in \eq{product}. Although this does not affect the conclusions reached in the next section on the counting of possible quantum states, it is unsatisfactory from the point of quantum simulation. In this section, we demonstrate how to recover a product formulae by making use of randomness. In Ref. \cite{WBHS10a1}, product formulae decompositions  \eq{product} were found for any local Hamiltonian, where the number of terms $N_p$ in the product depends on a smoothness parameter $\tilde \Lambda_P = \sup_{0 \le p \le P,0 \le t \le T}  \sum_X ( \| \partial ^p _t \tilde H_X(t) \|)^{1/(p+1)} $. In particular, these decompositions are inefficient when the fluctuation time-scale of the Hamiltonian becomes too small. Our methods circumvents these requirements by using randomness.

This is done in two steps. First, we can replace the time-ordered exponential integral with the exponential of an ordinary integral without introducing a significant error. Indeed, we show in Appendix \ref{A1} that
\begin{align}
&\left\|\cT \exp\left\{-i\int_{t_j}^{t_j+\Delta t}ds H_X(s)\right\} \right. \nonumber \\
&\quad\quad\left. - \exp\left\{-i\int_{t_j}^{t_j+\Delta t}ds H_X(s)\right\}\right\| \leq 2 \|H_X\|^2 \Delta t \label{eq:TnoT}
\end{align}
Using this result, we obtain the approximate decomposition
\begin{equation}
U(0,t) \approx \prod_{j=1}^n \prod_X \exp\left\{-i\int_{t_j}^{t_j+\Delta t} H_X(s)ds\right\}.
\label{eq:TOintegral}
\end{equation}
Note that this is still {\em not} a product formula because it involves integrals. This first step has nevertheless eliminated the need of a  time-order operator.

The second step to obtain a product formula for $U(0,t)$---one that does not require any integrals---makes use of randomness. The average Hamiltonian $H^{\rm Av}_{X,j}$ on the interval $[(j-1)\Delta t,j\Delta t]$ can be estimated using Monte Carlo integration. For every $j$, we can pick $m$ random times $\tau_j^k \in[t_j,t_j+\Delta t]$ and approximate $H^{\rm Av}_{X,j} \approx \frac 1m \sum_{k=1}^m H_X(\tau_j^k)$. Because the variance of the Hamiltonian is bounded by $\|H_X\|^2$, the sum converges to $\overline H^{\rm Av}_{X,j}$ with error estimate $\Delta t \|H_X\|/\sqrt m$. Using this Monte Carlo average, we can approximate the evolution operator of the time interval $[t_j,t_{j}+\Delta t]$ by
\begin{align}
U^{\rm Av}_X(t_j,t_j+\Delta t) &\approx \exp\{-i\frac 1m \sum_{k=1}^m H_X(\tau_j^k)\} \label{eq:app1}\\
&\approx \prod_{k=1}^m \exp\{-i\frac{\Delta t}{m} H_X(\tau_j^k)\}, \label{eq:app2}
\end{align}
where the order of the product can be chosen according to increasing values of $\tau_j^k$. The error in the first approximation \eq{app1} is set by the Monte Carlo estimate $\Delta t \|H_X\|/\sqrt m$ while the second approximation \eq{app2} is the usual Trotter-Suzuki. Summarizing, we can decompose the total evolution operator from time $0$ to $T$ as
\begin{equation}
U(0,T) \approx \prod_{j,k,X} \exp\{-i\frac{\Delta t}{m} H_X(\tau_j^k)\}
\label{eq:randomPF}
\end{equation}
where the product should be taken in increasing order of $\tau_j^k$ and any order of $X$.
This is a standard product formula \eq{product}---identical to the usual decomposition explained in the introduction \eq{simplePF} and the one presented in \cite{WBHS10a1}---except that the times $\tau_j^k$ at which the Hamiltonian is sampled are {\em random}. Thus, we see that by sampling the Hamiltonian at random times, we completely circumvent any smoothness requirements.

We note that this proof of \eq{TnoT} is an illustration of the decoupling principle  that tells us that the high-frequency fluctuations of the Hamiltonian should not affect the low-energy physics. As a consequence, it is possible to largely ignore these fluctuations---by replacing the time-dependent Hamiltonian by its average value on each time bin---without significantly modifying the dynamics of the system. This is the working principle behind renormalization group methods of quantum field theory and quantum many-body physics. The rotating wave approximation \cite{CTDRG92a} and effective Hamiltonian theory \cite{JJ07a} are  simple examples illustrating this principle in the case of time-dependent Hamiltonians.

More generally, we show in Appendix \ref{decoupling} that we can replace the time-dependent Hamiltonian $H(t)$ with a smoothed version $\tilde H(t)$ with fluctuation time-scale bounded by $\sigma$ without significantly affecting the resulting time-evolution operator. More precisely, we show that the time-evolution operators from time 0 to $T$ differ by at most $\|H\|^2 T \sigma$.

\medskip
\noindent{\em Counting states}---Let us now consider the set of all quantum states that can be reached starting from some fiducial state $\ket 0$ and evolving for some polynomial amount of time under any time-dependent Hamiltonian. A direct counting argument appears difficult because there are infinitely many distinct time-dependent Hamiltonians, and therefore there can {\em a priori} be infinitely many states in that set. However, we have just established that the time evolution operator generated by any one of these Hamiltonian can be well approximated by a polynomial-size quantum circuit.

Quantum circuits are discrete objects. Because the Hamiltonian is $k$-local, each term $\exp\{-i \Delta t \overline H_X(t_j)\}$ appearing in the product formula \eq{randomPF} is a $k$-body quantum gate. The Solovay-Kitaev theorem shows that each of these $k$-body gate can be approximated to accuracy $\epsilon$ by a product of at most $\log^c\frac 1\epsilon$ one- and two-qubit gates taken from a {\em finite discrete set} containing, say, $M$ gates, and where $c$ is a constant (close to 2).

Thus, to count the number of states that can be produced by arbitrary time-dependent Hamiltonians, it suffices to count the number of polynomial-size quantum circuits constructed from a universal discrete set of one- and two-qubit gates, and to consider an $\varepsilon$-ball around the output of each of these circuits, i.e. the set of states within a distance $\varepsilon$ of the outputs of these circuits. Surely, the states reached in polynomial time by arbitrary time-dependent Hamiltonians are contained in the union of these balls.

Since we limit the evolution to polynomial time, there exists a constant $\alpha$ such that the total number of gates in the simulation circuit is bounded by $K^\alpha$ where $K\propto N$ is the number of qubits required for the simulation. There are no more than $N_{\rm circuits} = (MK^2)^{K^\alpha}$ distinct ways of arranging these gates into a quantum circuit ($M$ possibility for each gate and $K^2$ possible pairs of qubits between which it can be applied), and therefore no more than $N_{\rm circuits}$ distinct states that can be produced. On the other hand, the states of $K$-qubits live on a $(2^{K+1}-1)$-dimensional hypersphere, whose surface area is $S = 2\pi^{2^{K}}/\Gamma(2^{K})$, and an $\varepsilon$-ball around a given state is a  $(2^{K+1}-2)$-dimensional hypersphere of volume $V = 2\pi^{2^K-1}\varepsilon^{2^{K+1}-2}/\Gamma(2^K)$.
Combining, we see that the $\varepsilon$-balls of physical states occupy only an exponentially small fraction $\frac{N_{\rm circuits}V}{S} = \cO(K^K \epsilon^{2^K})$ of the total volume in Hilbert space. Thus, we conclude that the overwhelming majority of states in the Hilbert space of a quantum many-body system can only be reached after a time scaling exponentially with the number of particles. For this reason, even for moderate size systems, these states are not physically accessible so they do not represent any state occurring in nature. For very large systems like in our expanding universe, it is indeed fair to assume that the time a system can evolve is not exponentially larger than the space it occupies.

\medskip
\noindent{\em Conclusion}---We have demonstrated that any time-dependent local Hamiltonian can efficiently be simulated using a quantum computer, independent of the frequencies involved. As an application, we have shown that the set of quantum states that can be reached from a product state with a polynomial time evolution of an arbitrary time-dependent quantum Hamiltonian is exponentially small. This means that the vast majority of quantum states in a many-body system are unphysical, as they cannot be reached in any reasonable time. As a consequence, all physical states live on a tiny submanifold, and that manifold can easily be parameterized by all poly-sized quantum circuits. This raises the question whether is makes sense to describe many-body quantum systems as vectors in a linear Hilbert space. The recent advances in real-space renormalization group methods \cite{CV09,VCM09,Vidal} indeed seem to suggest that a viable approach consists of parameterizing quantum many-body states using tensor networks and quantum circuits.

\medskip
\noindent{\em Acknowledgements}---This paper was presented at QIP 2011. We would like to thank I. Cirac and G. Vidal for inspiring discussions on the theme of this paper, and A. Winter for elucidating the counting argument that we used. We acknowledge funding of the ERC grant Querg, the European grant Quevadis and the FWF SFB grants Foqus and Vicom. DP is partially funded by NSERC and FQRNT.

%\bibliographystyle{/Users/dpoulin/archive/hsiam}
%\bibliographystyle{/Users/dpoulin/archive/qubib}
%\bibliography{qubib}

\appendix

\section{Average Hamiltonian}
\label{A1}

Define the average Hamiltonian for time bin
\begin{equation}
H^{\rm Av} = \frac 1{\Delta t} \int_{0}^{\Delta t} H(s) ds
\label{eq:av_H}
\end{equation}
and consider the time evolution operator $U^{\rm Av}(s) = \exp\{-is H^{\rm Av}\}$ that it generates. Our goal is to show that  $U^{\rm Av}(\Delta t)$ is close to  $U(\Delta t) = \cT \exp\left\{-i\int_0^s H(s) ds\right\}$ defined with a time-ordered integral.

Let $X(s) = \one - U^\dagger (s)U^{\rm Av}(s)$. We seek a bound on the norm
of $X(\Delta t)$. Since $X(0)=0$,  Schr\"odinger's equation yields
\begin{align*}
\label{X(t)}
X(\Delta t) &= \int_0^{\Delta t} ds \ \dot X(s)   -i \int_0^{\Delta t} dt \ U^\dagger (s)\Delta H(s) U^{\rm Av}(s)   \; ,
\end{align*}
with $\Delta H(s) = H(s) -  H^{\rm Av}$. Using the definition \eq{av_H}, this expression becomes
\begin{align*}
-\frac{i}{\Delta t} \int_0^{\Delta t}\int_0^{\Delta t} \left( U^\dagger (s)H(s) U^{\rm Av}(s)  -U^\dagger (s)H(\tau) U^{\rm Av}(s) \right) ds d\tau
\end{align*}
We can change the order of the integrals in the first term of this equation and use  the triangle inequality of the operator norm, to conclude that
\begin{align*}
\|X(\Delta t)\| &\leq \frac {\|H\| }{\Delta t}\int_0^{\Delta t}\int_0^{\Delta t} \left(\|U(s) - U(\tau)\| \right.\\
&+ \left. \|U^{\rm Av} (s) - U^{\rm Av} (\tau)\| \right) dsd\tau \\
&\leq 2 \|H\|^2 \Delta t
\end{align*}
where the last step folloows from Schr\"odinger's equation,
\begin{align}
\|U(s)-U(\tau)\| = \left\|\int_s^\tau H(t) dt\right\|.
\end{align}

\section{Decoupling principle}
\label{decoupling}

Suppose that we replace each term $H_X(t)$ in \eq H by its local time average, that is, we define $\tilde H_X(t) = \int G(t-s) H_X(s) ds$ where $G$ is a normal distribution of width $\sigma$ to be specified later, and we assume throughout that integrals are over the entire real axis unless otherwise stated.  If $T$ is the total evolution
time, we extend the definition of $H(t)$ so that $H(t)=0$ for $t<0$ or $t >T$.  Note that these integrals involve only $k$-body terms, so they can efficiently be carried-out classically \footnote{Although we want the Hamiltonian $H(t)$ to be as general as possible, we need to assume that these integrals exist. Piece-wise continuity is enough to ensure this. If these integrals do not exist only due to some short time intervals of total duration $\epsilon$, then these intervals can be left out of the simulation at the cost of an error at most $\epsilon \| H \|$. }. We similarly define $\tilde H(t) = \sum_X \tilde H_X(t)$. The idea behind the convolution
is to implement a low-pass filter with a frequency cutoff of order $\sigma^{-1}$.
 As we will now demonstrate, this has very little effect on the evolution of the system when $\sigma$ is sufficiently small.

Formally, we will compare the time-evolution operators $U$ and $\tilde U$ generated  by $H(t)$ and $\tilde H(t)$ respectively.  Let $X(t) = \one - U^\dagger(-\infty,t) \tilde U (-\infty,t)$. We seek a bound on the norm
of $X(\infty)$. Since $X(-\infty)=0$,  Schr\"odinger's equation yields
\begin{align}
X(\infty) = \int dt \ \dot X(t)   = -i \int dt \ U^\dagger(t) \Delta H(t) \tilde U(t)   \; ,
\end{align}
with $\Delta H(t) = H(t) - \tilde H(t)$.
By commuting the order of the integrals in $t$ and $s$ in Eq.~(\ref{X(t)}),
and using the fact that $G$ integrates to 1 and is symmetric,  we can bound $\|X(\infty) \|$ from above by
\begin{align}
\nonumber
 \int dt \int ds \;  G(t-s) \| U^\dagger(t) H(t) \tilde U(t) -
U^\dagger(s) H(t) \tilde U(s) \| \; .
\end{align}
Using the triangle inequality for the operator norm, we obtain
\begin{align}
\nonumber
\| X(\infty) \| \le   \|H \| \int_0^T dt \int  ds \; G(t-s) \times \\
\nonumber
\times [\| U^\dagger(t) - U^\dagger(s) \| +
\| \tilde U(t) - \tilde U (s) \| ] \; ,
\end{align}
where $\| H \| = \sup_t \|H(t) \|$ and we used the fact that $H(t)$ is
supported in $[0,T]$.
From Schr\"odinger's equation, it is easy to
show that $\| U^\dagger(t) - U^\dagger(s) \| \le |t-s| \|H \|$ and
$\| \tilde U(t) - \tilde U (s) \| \le |t-s| \| \tilde H \| \le |t-s| \| H \|$,
where the last step uses Young's inequality.
Combining these bounds, we obtain
\begin{align}
\nonumber
\| X ( \infty ) \| & \le 2 \| H \|^2 \int_0^T dt \int ds \; G(t-s) |t-s| \\
\label{result1}
& \le 2 \| H \|^2 T \sqrt{\frac 2 \pi} \sigma \; .
\end{align}
This is the desired result: choosing the width $\sigma$ of the normal distribution to be much smaller than $T \|H\|^2$ implies that the norm of $X(\infty)$---the difference between the exact evolution operator and the one obtained after truncating the high frequency components of $H(t)$---will be negligible.

The bound in Eq.~(\ref{result1}) is tight. To see this, consider the well known example of a two-level system with Hamiltonian
$H(t) =\lambda[ e^{i \omega t} \sigma^- + e^{-i \omega t} \sigma^+]$, where $\sigma^-$ ($\sigma^+$)
is the lowering (raising) Pauli operator. When $\omega \gg \lambda$
and $\omega T = 2 \pi k$, for
$k \in \mathbb{Z}$, the evolution $U(0,T)$ is well approximated by
$\exp(i T \lambda^2 \sigma_z/ \omega)$.
This is the so-called AC Stark shift. Averaging this Hamiltonian over a window of width $\sigma \gtrsim \frac 1\omega$ will truncate the unique oscillating terms, yielding $\tilde H(t) \approx 0$.
Therefore, $U(0,T) - \tilde U(0,T) \approx \exp(i T \lambda^2 \sigma_z/ \omega) - \one$,
and for short times $T$ the total approximation error  is of order $\lambda^2 T / \omega$, or
$\| H \|^2 T \sigma$, as claimed. This example and our general bound Eq.~(\ref{result1}) reflect the fact
that high-frequency terms can only affect the overall
evolution at second order in perturbation theory.

\end{document}